\documentclass[preprint,showpacs,preprintnumbers,amsmath,amssymb,nofootinbib]{revtex4}
\usepackage{epsfig}
\usepackage{epstopdf}
\usepackage{bbm}
\usepackage{color}
\newcommand{\be}{\begin{equation}}
\newcommand{\ee}{\end{equation}}
\newcommand{\bea}{\begin{eqnarray}}
\newcommand{\eea}{\end{eqnarray}}

\def\simlt{\mathrel{\raise.3ex\hbox{$<$\kern-.75em\lower1ex\hbox{$\sim$}}}}
\def\simgt{\mathrel{\raise.3ex\hbox{$>$\kern-.75em\lower1ex\hbox{$\sim$}}}}

\usepackage{scalerel}[2014/03/10]
\usepackage{stackengine}
\newcommand\reallywidehat[1]{\ThisStyle{%
  \setbox0=\hbox{$\SavedStyle#1$}%
  \stackengine{-1.0\ht0+.5pt}{$\SavedStyle#1$}{%
    \stretchto{\scaleto{\SavedStyle\mkern.15mu\char'136}{2.6\wd0}}{1.4\ht0}%
  }{O}{c}{F}{T}{S}%
}}

\begin{document}
%

\title{Higgsino Dark Matter in a Non-Standard History of the Universe}

\author{Chengcheng Han$^1$ \\~ \vspace*{-0.4cm}}
\affiliation{
$^1$ School of Physics, KIAS, 85 Hoegiro, Seoul 02455, Republic of Korea
  \vspace*{1.5cm}
}

\begin{abstract}
A light higgsino is strongly favored by the naturalness, while as a dark matter candidate it is usually under-abundant. We consider the higgsino production in a non-standard history of the universe, caused by a scalar field with an initially displaced vacuum. We find that given a proper reheating temperature induced by the scalar decay, a light higgsino could provide the correct dark matter relic abundance. On the other hand, a sub-TeV higgsino dark matter, once observed, would be a strong hint of the non-standard thermal history of the universe.
\end{abstract}

\maketitle

\section{Introduction}
Although supersymmetry has evaded all observations,  it is still well motivated for solving the naturalness problem as well as providing a WIMP dark matter candidate. In the minimal supersymmetric standard model(MSSM), assuming the conservation of the R-parity, usually the lightest neutralino plays the role of dark matter. It is a mixing state of the higgsino and gauginos, whose mass is determined by the higgsino mass parameter $\mu$ and gaugino mass parameters $M_{1,2}$.  At the same time, the $\mu$ term is also involved in the Higgs mass matrix.  If it is much beyond a hundred GeV, a fine-tuning would be present  in the Higgs mass. Such a fine-tuning, not strictly, can be roughly estimated by 
\begin{eqnarray}
\Delta_{min}\simeq \frac{2\mu^2}{M_h^2}
\end{eqnarray}
To avoid too much fine-tuning in the Higgs mass, a moderate higgsino mass parameter $\mu$ around hundreds of GeV is preferred. However, it is known that a sub-TeV higgsino-like neutralino is under-abundant, while a bino-like neutralino generally requires the $\mu$ parameter lager than TeV to evade the dark matter direct detection \cite{Badziak:2017the, Han:2016gvr} \footnote{We note that there exists still some blind spot regions \cite{Cheung:2012qy, Huang:2014xua} where $\tan \beta$ is tuned to satisfy $\mu \sin 2\beta + M_1=0$. Here we do not focus on such special region.}. These observations force us to accept the fact that either MSSM only provides part of the dark matter in our universe or it has been fine-tuned at least around $\mathcal{O}(200)$.

On the other hand, for the WIMP dark matter, one often postulates that the dark matter freezes out in a radiation-dominated universe. The thermal freeze-out temperature is around  $m_\chi/20$ which is typically a  few GeV to tens of GeV for a WIMP dark matter.  However,  since we have no information about the details of the early universe before the big bang nucleosynthesis(BBN), there is no reason to assume the universe must be radiation-dominated when dark matter freeze-out happens. One of the variations is that there may exist a matter-dominated epoch before the BBN, caused by a scalar field with an initially displaced vacuum and undergoing a coherent oscillation when the Hubble parameter becomes smaller than the mass of the scalar. Such a scalar is ubiquitous in new physics models, for instance, the axion-like particles(ALPs) \cite{Weinberg:1977ma, Wilczek:1977pj}, the moduli \cite{Coughlan:1983ci,Banks:1993en,  deCarlos:1993wie}, or the flaton \cite{Lyth:1995ka}, etc.. If the dark matter freezes out in a non-standard history of the universe, the relic density of the dark matter would be much different. Such a scenario is well studied in the literature \cite{McDonald:1989jd, Chung:1998rq, Moroi:1999zb, Giudice:2000ex, Kamionkowski:1990ni, Jeannerot:1999yn, Lin:2000qq, Moroi:1994rs, Kawasaki:1995cy, Allahverdi:2002nb, Allahverdi:2002pu, Khalil:2002mu, TorrenteLujan:2002ss, Fornengo:2002db, Pallis:2004yy, Endo:2006zj, Nakamura:2006uc, Hashimoto:1998mu, Kohri:2004qu, Kohri:2005ru, Gelmini:2006pw, Gelmini:2006pq, Acharya:2009zt, Arcadi:2011ev}, and emphasized recently by \cite{Drees:2017iod,Drees:2018dsj,Kane:2015jia,Arbey:2018uho,Arias:2019uol, Aparicio:2016qqb, Harigaya:2013vwa, Harigaya:2014waa, Mukaida:2015ria, Hardy:2018bph, Hamdan:2017psw, Maldonado:2019qmp}.

In this paper, concerning the naturalness, we study the higgsino dark matter production in a non-standard thermal history of the universe. We will show that a light higgsino could be a natural dark matter candidate with the correct relic abundance \footnote{There are also other possibilities, for example, see \cite{Barenboim:2014kka, Bae:2013hma, Choi:2008zq}. }. The paper is organized as follows. We first overview the calculation of the dark matter relic density in the matter-dominated universe, and then we show how the higgsino dark matter could be accommodated in such a framework. In the end, we discuss the detections of the higgsino dark matter.

\section{higgsino Dark Matter in an early matter-dominated universe}
It is well-believed that the early universe undergoes a rapid expansion period which is called inflation, after which the universe is reheated and the radiation-dominated epoch starts with an initial temperature. The radiation-dominated epoch continues to a very late stage until the matter component of the universe takes over. This is the common lore of the history of our universe. However, if there exists a scalar initially displaced from the minima of the vacuum, when the Hubble parameter becomes smaller than the scalar mass, it will start to oscillate around the minima and behave as a matter component of the universe. It soon dominates the universe and the universe becomes matter-dominated. After the decay of the scalar, the universe is reheated \footnote{If the scalar does not decay, it could be a dark matter candidate.} with a temperature at least larger than 4 MeV without destroying the BBN \cite{Ellis:1990nb, Kawasaki:2017bqm, Forestell:2018txr, Hufnagel:2018bjp}. We denote such a scalar as $\phi$, with mass $m_\phi$, decay width $\Gamma_\phi$ and the initial displaced value $\phi_I$(for the moduli $\phi_I \sim M_{Pl}$). The decay width can be calculated by
\begin{eqnarray}
\Gamma_\phi \simeq \frac{C}{8\pi} \frac{m_\phi^3}{\Lambda^2}
\end{eqnarray}
where $\Lambda$ is the typical scale where the scalar couples to the standard model sector (for the moduli, $\Lambda=M_{Pl}$). $C$ is an $\mathcal{O}(1)$ parameter depending on the detail of the physics.
 The reheating temperature can be defined when the life time $1/\Gamma_\phi$ is equal to the time scale of the universe $1/H$:
\begin{eqnarray}
T_{rh} \equiv   (90/\pi^2 g_*)^{1/4} \sqrt{M_{Pl}\Gamma_\phi}.
\end{eqnarray}
During this epoch, the related Boltzmann equation can be written as \cite{Drees:2017iod}:
\begin{eqnarray}
\frac{d\rho_\phi}{dt}&=&- 3H \rho_\phi-\Gamma_\phi \rho_\phi \\
\frac{d\rho_r}{dt}&=&- 4H \rho_r+\Gamma_\phi \rho_r(1-\frac{b \langle E_\chi \rangle}{m_\phi}) +  \langle \sigma v \rangle  2  \langle E_\chi \rangle [ n_\chi^2- (n_\chi^{eq})^2] 
\label{rhor} \\
\frac{dn_\chi}{dt}&=&- 3H n_\chi+\frac{b}{m_\phi} \Gamma_\phi \rho_\phi-  \langle \sigma v \rangle    [ n_\chi^2- (n_\chi^{eq})^2]  
\label{dm}
\end{eqnarray}
where $\rho_\phi, \rho_r$ are the energy density of the scalar field and radiation respectively and $n_\chi$ is the number density of the dark matter.  $\langle E_\chi \rangle \simeq \sqrt{m_\chi^2+3 T^2}$ is the average energy of the dark matter. The second term of right side of Eq. (\ref{dm}) denotes the dark matter non-thermal production from the scalar decay while the third term is the dark matter annihilation. Here $b$ can be understood to be the branching ratio of the scalar decaying into the higgsino sector, and $\langle \sigma v \rangle$ should be taken as
\begin{eqnarray}
\langle \sigma v \rangle &=& \frac{\sum^N_{i,j} w_i w_j \sigma_{ij} x^{-n}}{(\sum_i w_i)^2} \label{sigmav} \\
w_i&=&(\frac{m_{\chi_i}}{m_\chi})^{3/2} e^{-x(m_{\chi_i}/m_\chi -1)}
\end{eqnarray}
where $x = m_\chi/T$ and  all the higgsino components $(\tilde{H}^0_1, \tilde{H}^0_2, \tilde{H}^+, \tilde{H}^- )$ should be included in the calculation in Eq. (\ref{sigmav}) .
The temperature $T$ and entropy $s$ can be derived as 
\begin{eqnarray}
T&=& \left ( \frac{30 \rho_r}{ \pi^2 g_*} \right )^{1/4}  \label{radiation}\\
s&=&\frac{2\pi^2}{45} h_* T^3 \label{entropy}
\end{eqnarray}
To solve the Boltzmann equation, we define the following dimensionless parameters:
\begin{eqnarray}
\Phi \equiv \rho_\phi T_{rh}^{-1} a^3;  R \equiv \rho_r a^4 ;  X \equiv n_\chi a^3;  A \equiv a T_{rh}
\end{eqnarray}
The boltzmann equation can be rewritten as \cite{Drees:2017iod}
\begin{eqnarray}
 \tilde{H}\frac{d\phi}{dA}&=& - c_\rho^{1/2} A^{1/2} \Phi  \\
 \tilde{H}\frac{dR}{dA}&=& c_\rho^{1/2} A^{3/2} (1-\frac{b \langle E_\chi \rangle}{m_\phi}) \Phi   
+c^{1/2}_1 A^{-3/2}  \langle \sigma v \rangle  2\langle E_\chi \rangle M_{Pl} (X^2-X_{eq}^2)   \label{R} \\
 \tilde{H}\frac{dX}{dA}&=&  A^{1/2} T_{rh} \Phi \frac{b}{m_\phi}   
  -c^{1/2}_1 A^{-5/2} \langle \sigma v \rangle M_{Pl} T_{rh} (X^2-X_{eq}^2)
 \end{eqnarray}
where $c_\rho= \frac{\pi^2 g_*(T_{rh})}{30}$, $c_1= \frac{3}{8\pi}$ and 
\begin{eqnarray}
 \tilde{H}\equiv \sqrt{ \Phi + R/A+ X \langle E_\chi \rangle / T_{rh}} 
\end{eqnarray}
The Hubble parameter $H$ is 
\begin{eqnarray}
H=\tilde{H} T^2_{rh} A^{-3/2} c_1^{-1/2}/M_{Pl} 
\end{eqnarray}
In the hot universe, instead of using $R$ in Eq. (\ref{R}) it is more convenient to use the temperature $T$ as a parameter 
\begin{eqnarray}
\frac{dT}{dA}= (1+ \frac{1}{3} \frac{d \log h_{*} }{d \log T})^{-1} \left [  -\frac{T}{A}+ \frac{\Gamma_\phi \rho_\phi}{3 H s A}(1-\frac{b \langle E_\chi \rangle}{m_\phi}) + 
\frac{2  \langle E_\chi \rangle  \langle \sigma v \rangle T^6_{rh} }{3 H s A^7}(X^2-X_{eq}^2)           \right ]
\end{eqnarray}
The energy density of radiation and entropy can be related in Eq.~(\ref{radiation}) and Eq.~(\ref{entropy}).
We note that the factor $\frac{1}{3} \frac{d \log h_{*} }{d \log T}$ could be as large as $\sim 0.4$ when the temperature get close to the QCD confinement scale.  It is particularly important when $T_{rh}$ becomes close or below this scale. 

Assuming $b \ll 1$, the temperature during the matter-dominated epoch can be estimated as  \cite{Giudice:2000ex}
\begin{eqnarray}
T\simeq T_{\rm max} A^{-3/8}
\end{eqnarray}
where $T_{\rm max} \simeq M^{1/4}_{Pl} H_I^{1/4} T_{rh}^{1/2}$ ($H_I= \sqrt{\frac{4\pi m_\phi^2 \phi_I^2}{3 M^2_{Pl}}}$).  During the matter-dominated epoch, the Hubble parameter $H$ is around $\frac{T^4}{T^2_{rh}  M_{Pl}}$ comparing with $H \simeq \frac{T^2}{M_{Pl}}$ in the radiation-dominated universe. Depending on the $T_{rh}$ and dark matter freeze-out temperature  $T_{fo}$, the results can be summarized as follows.

\begin{figure}[ht]
\centering
\includegraphics[width=3.5in]{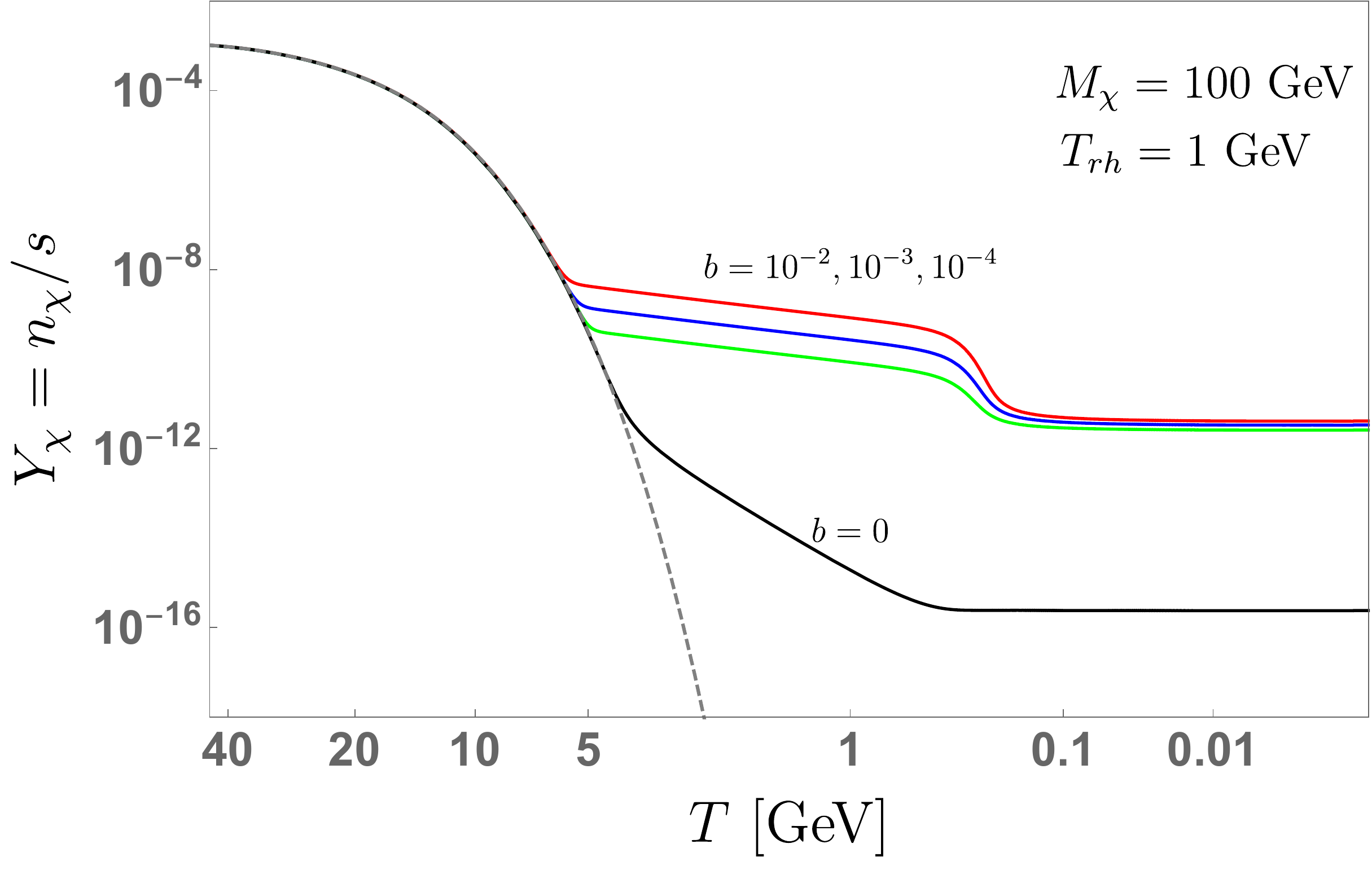}
\caption{Illustration of the dark matter relic abundance with different $b$. The black, red, blue, green line show the evolution of $Y_\chi$ with $b= 0, 10^{-2}, 10^{-3}, 10^{-4}$ respectively. The higgsino mass is fixed to be 100 GeV and the reheating temperature is set to be 1 GeV. Here the scalar mass $m_\phi=10$ TeV.}
\label{fig1}
\end{figure}

\begin{enumerate}
\item $T_{rh} \gg T_{fo}$: in this case, the dark matter relic density is the same as the usual freeze-out in radiation epoch. Since we are considering a higgsino dark matter mass around a few hundred GeV, this case is not so interesting to us.

\item $T_{rh} \ll T_{fo}$ and $b=0$: in this case, the dark matter freezes out when the universe is still matter-dominated. As shown with the black line in the Fig. \ref{fig1}, for the 100 GeV higgsino dark matter, the freeze-out temperature is around 5 GeV $\sim m_\chi/20$. After the dark matter freezes out, the comoving number density of the dark matter remains a constant and thus $n_\chi \propto A^{-3}$, while the entropy $s$ is proportional to   $A^{-9/8}(A^{-3})$ in the matter-dominated(radiation-dominated) epoch, hence $Y_\chi  \propto A^{-15/8}(A^0) $ in the matter-dominated(radiation-dominated) epoch. The final relic density can be approximated by \cite{Giudice:2000ex}
\begin{eqnarray}
\Omega h^2= \Omega_0 h^2     \frac{T_{fo}}{T^\prime_{fo}}  \left( \frac{T_{rh} }{ T^\prime_{fo} } \right )^3
\label{relic1}
\end{eqnarray}
where $T^\prime_{fo}$ is the freeze-out temperature in the matter-dominated epoch, it is above but not too much larger than $T_{fo}$, and $\Omega_0 h^2 $ is the relic density calculated in the normal freeze-out where a radiation-dominated universe is assumed. As shown in Eq. (\ref{relic1}), the dark matter relic abundance becomes even smaller. The main reason is that dark matter number density is diluted by the entropy release from the scalar decay. In this case, again for a higgsino mass below 1 TeV, it can only provide part of the dark matter in our universe.

\item $T_{rh} \lesssim c T_{fo}$ ($c \sim \mathcal O(4)$) and $b\ne0$: in this case, the dark matter is mainly non-thermally produced from the scalar decay. The dark matter starts to follow the thermal equilibrium, and soon it freezes out when the production rate from the scalar decay becomes equal to the annihilation rate. At this stage, 
\begin{eqnarray}
n_\chi \approx  \sqrt{      \frac{ T^2_{rh} b \rho_\phi} { M_{Pl} m_\phi  \langle \sigma v \rangle}  } \propto A^{-3/2}
\end{eqnarray}
Considering $s \propto A^{-9/8}$ in matter-dominated epoch, we get $Y_\chi=n_\chi/s \propto A^{-3/8}$. In Fig. \ref{fig1} it clearly shows the evolution of $Y_\chi$ for $b= 10^{-2}, 10^{-3},10^{-4}$ is flatter than that in the case $b=0$. More interestingly, in addition to the freeze-out, there is a re-annihilation process at a temperature $T\simeq T_{rh}$. The main reason is that after the total decay of the scalar, the dark matter number density is too large that the dark matter annihilation rate becomes much larger than the expansion rate of the universe, i.e. $n_\chi(T_{rh}) { \langle \sigma v \rangle} \gg {H(T_{rh})}$. Then the dark matter start to re-annihilate until $n_\chi(T_{rh}) \simeq \frac{H(T_{rh})}{ \langle \sigma v \rangle}$ satisfied.  Note that the final dark matter number density after re-annihilation is not sensitive to the value of $b$. As shown in Fig. \ref{fig1}, even if the $b=10^{-2},10^{-3},10^{-4}$, the final dark matter relic density could be very similar. Comparing to the normal freeze-out, the relic density of the dark matter can be approximated by \cite{Giudice:2000ex}
\begin{eqnarray}
\Omega h^2= \Omega_0 h^2     \frac{T_{fo}}{T_{rh}}  
\label{omega}
\end{eqnarray}
This result is independent of $b$, only $T_{rh}$ involved. In the literature this scenario is often denoted as ``non-thermal WIMP miracle" \cite{Acharya:2009zt} or ``re-annihilation  scenario" \cite{Gelmini:2006pw}. In this case, the dark matter relic abundance is enhanced with a factor  $\frac{T_{fo}}{T_{rh}} $. Therefore, with a given $T_{rh}$ one can always find a light higgsino with a correct dark matter relic abundance. \footnote{Note there is another scenario where $b$ is very small and the re-annihilation never happens. We do not study this scenario here due to the following reason: in supersymmetry, if the scalar can decay into SM sector, it can also decay into the SUSY sector. We expect such decay is also sizable in SUSY sector. To satisfy $n_\chi(T_{rh}) < \frac{H(T_{rh})}{ \langle \sigma v \rangle}$ where re-annihilation never happens, we need a rather small $b$ ($b  \lesssim 10^{-7}$ for our benchmark point).}
\end{enumerate}
We note that even if the reheating temperature is defined as the typical temperature at which scalar decays, only around 20\% of the energy density of $\phi$ is transferred into the radiation at $T=T_{rh}$. The total decay completes at a later time, that is why the re-annihilation happens at a lower temperature than $T_{rh}$.

Finally the dark matter relic density can be calculated by
\begin{eqnarray}
\Omega_{dm}h^2=\frac{Y_{\infty} s_0 m_\chi}{\rho_c/h^2} 
\end{eqnarray}
where the current entropy $s_0=2.89\times 10^9$ m$^{-3}$, $M_{Pl}=1.22\times 10^{19}$ GeV, $\rho_c/h^2=10.5$ GeVm$^{-3}$.

\begin{figure}[ht]
\centering
\includegraphics[width=3.5in]{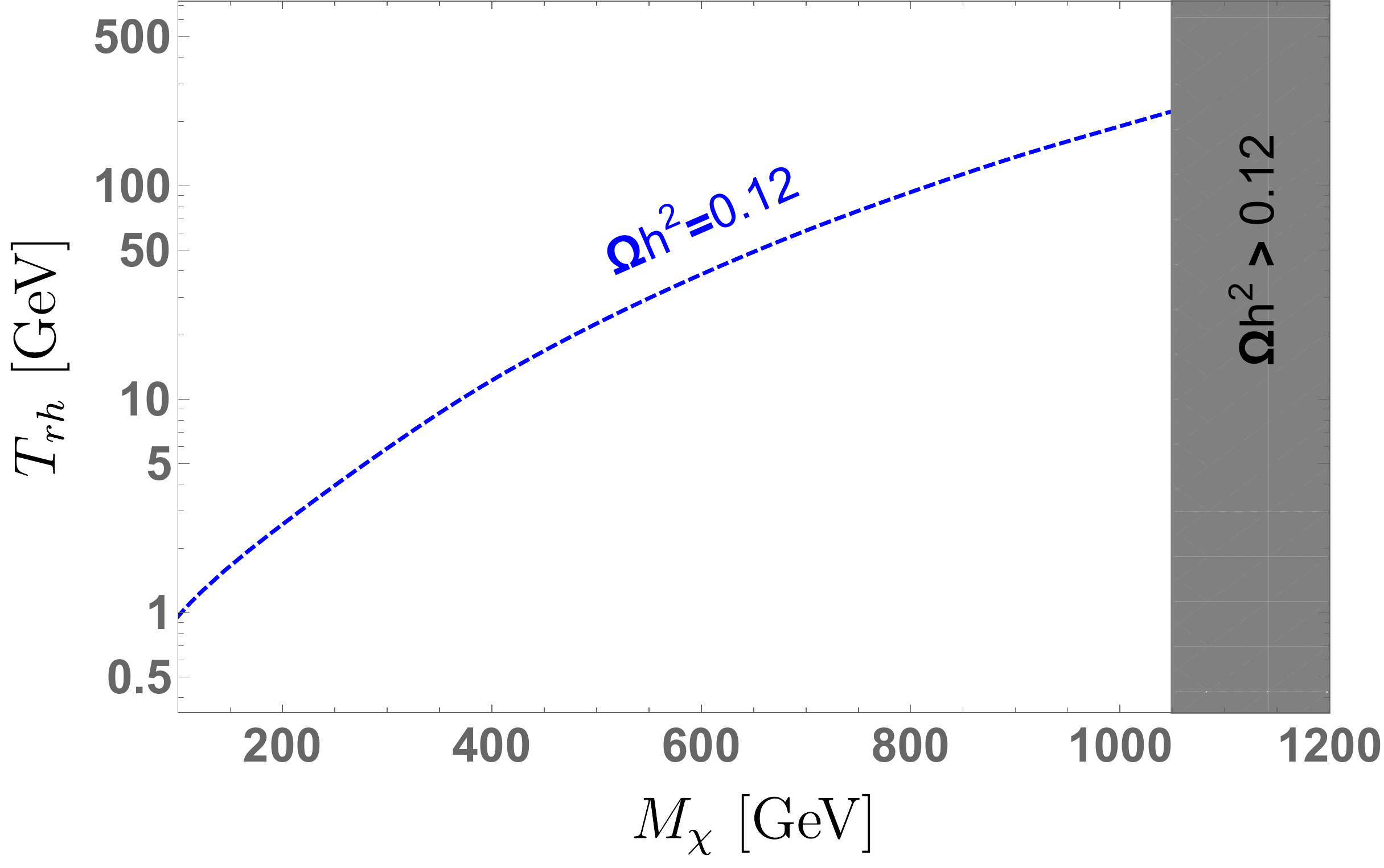}
\caption{The required reheating temperature versus higgsino dark matter mass. The blue dashed line is the reheating temperature $T_{rh}$ which makes the higgsino satisfy the correct dark matter relic density. For a higgsino mass beyond 1 TeV, it is usually over-abundant. }
\label{fig2}
\end{figure}

Since in the re-annihilation scenario the final number density of dark matter is not sensitive to the value of $b$, in the following calculation we fix $b=10^{-2}$.  Our numerical results are checked by comparing the dark matter relic density calculated in the limit  $T_{rh} \gg T_{fo}$ (in practice $T_{rh} =\frac{m_\chi}{2}$) with the results calculated by {\sc MicrOMEGAs} \cite{Belanger:2006is}  which assumes dark matter freeze-out during the radiation era. We find that these two results are well consistent with each other and the deviation is within a few percents. In our calculation, except for the higgsino mass parameter $\mu$,  we fixed all the other SUSY mass parameters to be 3 TeV. The $A_t$ and $\tan\beta$ are set to be 4 TeV and 20 respectively to fit a Higgs mass around 125 GeV. 

In principle, we can always get the correct dark matter relic abundance for a light higgsino through tuning the reheating temperature $T_{rh}$\footnote{Although such possibility  is pointed by \cite{Allahverdi:2012wb}, here without using the approximation Eq. (\ref{omega}), we performed more detail numerical studies.}. In Fig. \ref{fig2} the dashed blue line tells the required $T_{rh}$ versus the corresponding higgsino mass. It shows that the required reheating temperature could vary from 1 GeV to 230 GeV with the corresponding higgsino mass from 100 GeV- 1 TeV.  The typical reheating temperature for 100 GeV higgsino is around 1 GeV to satisfy the correct dark matter relic density. For a higgsino mass beyond 1 TeV, the higgsino dark matter is usually over-abundant. Actually in that case, if we set $b=0$ and a proper $T_{rh}$, heavy higgsino dark matter is also applicable. Since we are only interested in a light higgsino, here we do not present the related results. On the other hand, if we finally find a higgsino dark matter below 1 TeV, we can easily refer to the reheating temperature from Fig. \ref{fig2}. 

In the following, we discuss more details about the detection of the higgsino dark matter \footnote{It is also possible to probe such scenario by gravitational wave searches \cite{DEramo:2019tit}.}.

\begin{figure}[ht]
\centering
\includegraphics[width=3.0in]{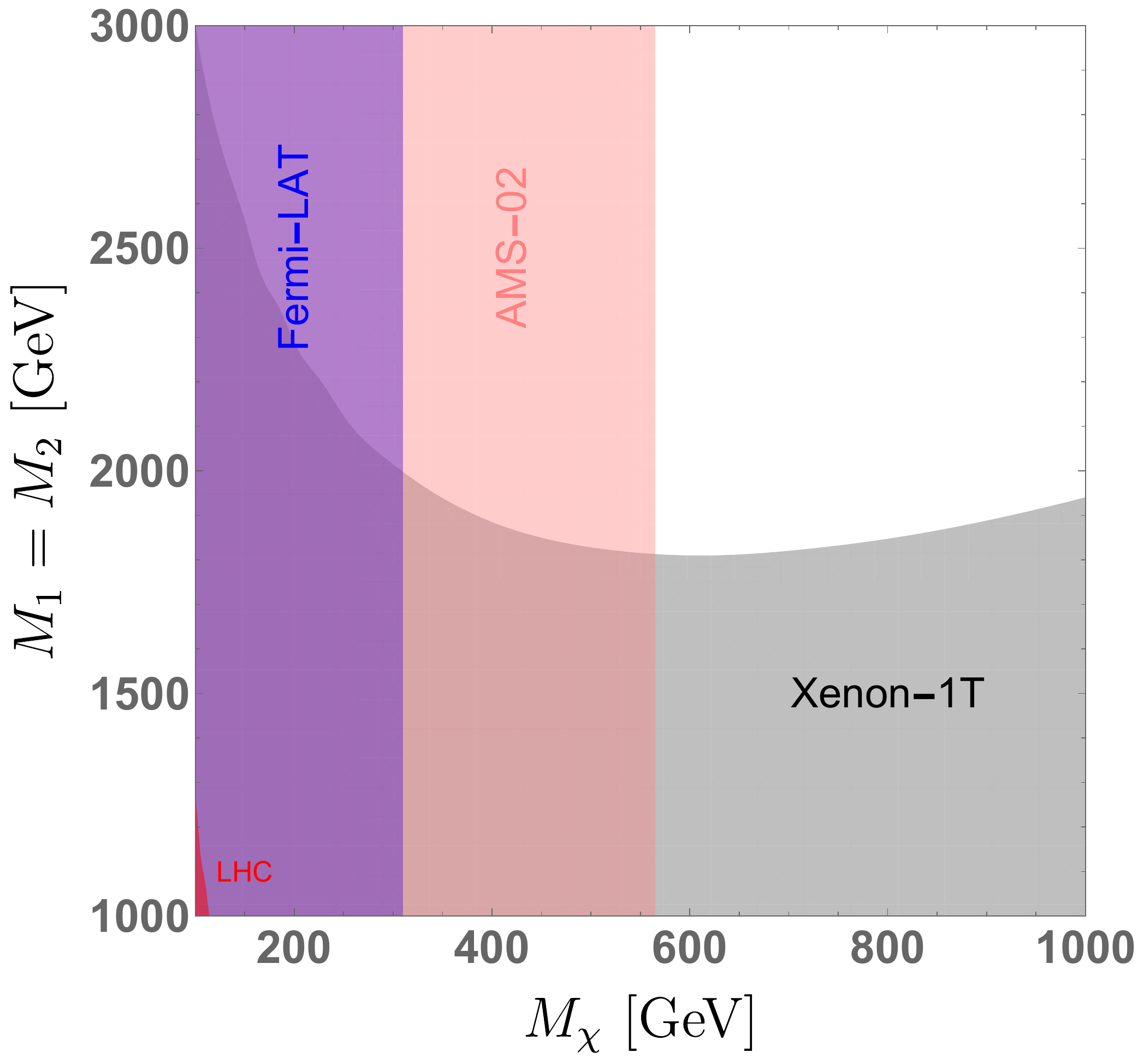}
\caption{Current limit on the higgsino dark matter. The gray region is from dark matter direct searches Xenon-1T for one-year observation. The light blue region is limit from the dwarf galaxy $\gamma$-ray  searches at Fermi-LAT. The pink region is excluded by the AMS-02 antiproton data. The red region is from LHC searches for soft leptons plus missing energy.}
\label{fig3}
\end{figure}

\subsection{Direct searches for the higgsino DM}
In MSSM, the dark matter-nucleon interactions are mainly mediated by the Higgs boson. For a higgsino dark matter, the higgsino-higgsino-Higgs coupling origins from the higgsino-Higgs-gaugino coupling with mixing between gaugino and higgsino. In Fig. \ref{fig3} the gray region is covered by the dark matter direct search from Xenon-1T \cite{Aprile:2018dbl}, where we fix $\tan \beta=20$ and $M_1=M_2$ with varying $M_1$ from 1 TeV to 3 TeV.  It shows a universal gaugino mass below 1.8 TeV has been excluded. 

\subsection{Indirect searches for the higgsino DM}
For the higgsino dark matter, at present it still annihilates into SM particles and provides an astronomical signal. The major annihilation channels for higgsino dark matter are $W^+W^-$ and $ZZ$, which would lead to a gamma ray signal in the galaxies. Here we consider the Fermi-LAT gamma ray observations from dwarf-spheroidal galaxies \cite{Fermi-LAT:2016uux}. The $\gamma$-ray flux from dark matter annihilation is 
\begin{eqnarray}
\phi(\Delta \Omega, E_{\rm min}, E_{\rm max})=\frac{1}{4\pi} \frac{\langle \sigma v \rangle}{2 m^2_{\rm DM}} \int^{E_{\rm max}}_{E_{\rm min}} \frac{d N_\gamma}{d E_{\gamma}} dE_{\gamma} \times J_{\rm factor}
\end{eqnarray}
In our numerical calculation, the gamma spectrum $\frac{d N_\gamma}{d E_{\gamma}}$ is simulated by {\sc Pythia-8.2} \cite{Sjostrand:2014zea} and the likelihood for each dwarf galaxy is calculated by \cite{Ackermann:2015zua}:
\begin{eqnarray}
\mathcal L (m_\chi, \langle \sigma v \rangle , J ) &=& \mathcal L(J |  J_{\rm obs} , \sigma_J) \prod^{24}_{j=1} \mathcal L_j( \Phi_j (m_\chi, \langle \sigma v \rangle, J)), \\
\mathcal L(J |  J_{\rm obs} , \sigma_J) &=& \frac{1}{\log(10) J_{\rm obs} \sqrt{2\pi} \sigma_J} e^{-(\log_{10} J-\log_{10} J_{\rm obs})^2/ 2 \sigma_J^2}
\end{eqnarray}
The 24 bins of $\mathcal L_j( \Phi_j (m_\chi, \langle \sigma v \rangle, J))$ are calculated by using the tabulated likelihood provided by Fermi-LAT for each dwarf galaxy and energy flux. The limit on $ \langle \sigma v \rangle$ is derived by requiring 
\begin{eqnarray}
-2 \Delta\log \mathcal L (m_\chi, \langle \sigma v \rangle) \equiv -2 \log \frac{\mathcal L (m_\chi, \langle \sigma v \rangle , \hat {\hat J} )}{\mathcal L (m_\chi, \reallywidehat {\langle {\sigma v} \rangle} , {\hat J} )} \leq 2.71
\end{eqnarray}
where $\hat {\hat J}$ is the $J$-factor that maximizes the likelihood for a given $\langle \sigma v \rangle$ and $m_\chi$, while the $\mathcal L (m_\chi, \reallywidehat {\langle {\sigma v} \rangle} , {\hat J})$ is the global maximum of the likelihood for a given $m_\chi$. Instead of combining all the dwarf galaxy analysis, we just simplify to adopt the strongest limit on the $\langle \sigma v \rangle$ for all the dwarf galaxy analysis. Finally, we find a higgsino dark matter with a mass less than 310 GeV has been excluded by the gamma-ray searches \footnote{A naive interpolation of the combined results of MAGIC and Fermi-LAT on $\chi \chi \rightarrow W^+W^-$ channel \cite{Ahnen:2016qkx} exclude a higgsino mass less than 350 GeV. }.

Since the higgsino dark matter dominantly annihilates into $W^+W^-/ZZ$ which mainly decays into hadronic objects, the AMS-02 antiproton data could give a strong limit on higgsino dark matter. It is shown~\cite{Cuoco:2017iax} that a higgsino dark matter less than 565 GeV has been excluded. Even considering a conservative estimate including the systematic uncertainties, the limit is around 330 GeV which is comparable with the Fermi-LAT gamma-ray constraint. In Fig. \ref{fig3} we draw the higgsino dark matter limit from AMS-02 antiproton data in the pink region without including the systematic uncertainties.

\subsection{Collider searches for the higgsino DM}
It is known that the collider search for higgsino is very difficult. The major difficulty is due to the degeneracy of the spectrum. For $|M_1|$, $|M_2| \gg \mu$,  the mass splitting
\begin{eqnarray}
\Delta m \equiv m_{\tilde \chi^0_2}- m_{\tilde \chi^0_1} \approx \frac{m_W^2}{M_2} +\frac{m_W^2 \tan^2 \theta_W}{M_1}
\end{eqnarray}
For the chargino the mass splitting is even smaller, $m_{\tilde \chi^\pm_1}- m_{\tilde \chi^0_1} \approx \frac{\Delta m}{2}$. Depending on the gaugino mass, the mass difference can vary from sub-GeV to 30 GeV. For the mass splitting around sub-GeV, the only way to look for such higgsino is through monojet plus missing energy at the large hadron collider(LHC). With high luminosity LHC, it is possible to probe higgsino below 150 GeV \cite{Han:2013usa}. For a larger mass splitting, the soft leptons from the heavier neutral higgsino can be tagged and taken to be a typical signal of higgsino \cite{Han:2014kaa, Baer:2014kya, Han:2015lma}. LHC already performed such searches \cite{Aaboud:2017leg} and give a limit around 100-150 GeV with $\Delta m$ varying from 2 GeV- 20 GeV. We add such a limit in Fig. \ref{fig3}, and it shows only a tiny region is covered by the current LHC, while this region is readily excluded by the other dark matter searches. Other ways to look for the higgsino can be found in \cite{Fukuda:2017jmk, Matsumoto:2017vfu, Han:2018rkz, Schwaller:2013baa, Curtin:2017bxr, Mahbubani:2017gjh}.

\subsection{Higgsino DM searches from BBN and CMB}

Even if the higgsino dark matter freezes out at the temperature around $m_\chi/20$, it continues annihilating in a later time, resulting in some observable effects from the early universe.  
Firstly the higgsino dark matter annihilation produces energetic hadrons during the big-bang nucleosynthesis(BBN) epoch, which affects the abundance of the light-elements. Such effect has been well studied in the literatures~\cite{Reno:1987qw, Frieman:1989fx, Jedamzik:2004ip, Hisano:2011dc, Henning:2012rm, Kawasaki:2015yya} and~\cite{Kawasaki:2015yya} shows that a limit on the dark matter annihilation cross section(only $W^+/W^-$ channel) is around  $2\times 10^{-25} {cm}^3/s$ for a 100 GeV dark matter and $2\times 10^{-24} {cm}^3/s$ for a 300 GeV dark matter. Such bound excludes a higgsino dark matter around 200 GeV. On the other hand, the annihilation of the higgsino dark matter also injects extra energy in the recombination epoch, changing the ionization fraction of the neutral hydrogen. The measurements of the CMB anisotropy is sensitive to such energy injection and set a limit on the dark matter annihilation cross section~\cite{Galli:2009zc, Cirelli:2009bb, Slatyer:2009yq, Lopez-Honorez:2013lcm, Kawasaki:2015peu}. It is shown~\cite{Kawasaki:2015peu} that such limit is comparable with the BBN.  In Fig. \ref{fig3} we do not add these limits because they are much weaker than the limit from dark matter indirect searches.

\section{Summary and Conclusions}
We consider the higgsino DM production in a non-standard thermal history of the universe. Given a proper the reheating temperature,  we find a light higgsino could provide the correct dark matter relic abundance.  However, the AMS-02 antiproton data already excluded a higgsino mass below 565 GeV. One the other hand, the discovery of the light higgsino dark matter below 1 TeV would be a strong hint of the non-standard thermal history of the universe. A reheating temperature can be referred to depending on the mass of the higgsino dark matter.

\section*{Acknowledgements}
C. Han thank Jongkuk Kim, Feng Luo, Eung Jin Chun, Yi-Lei Tang, Peiwen Wu, Jin-Min Yang for helpful discussions and useful suggestions.

\end{document}